\documentclass{aa}  
\usepackage{graphicx}
\usepackage{txfonts}
\usepackage[figuresright]{rotating}
\begin{document}
\title{Low-resolution spectroscopy and spectral energy distributions \\ 
of selected sources towards $\sigma$~Orionis}
\titlerunning{Selected sources towards $\sigma$~Orionis}
%
%\subtitle{} 
%
\author{J. A. Caballero\inst{1}
\and
L. Valdivielso\inst{2}
\and
E. L. Mart\'{\i}n\inst{2,3}
\and
D. Montes\inst{1}
\and
S. Pascual\inst{1}
\and
P. G. P\'erez-Gonz\'alez\inst{1}
}
\offprints{Jos\'e A. Caballero, investigador Juan de la Cierva at the UCM
(\email{caballero@astrax.fis.ucm.es}).}  
\institute{
Dpto. de Astrof\'{\i}sica y Ciencias de la Atm\'osfera, Facultad de F\'{\i}sica,
Universidad Complutense de Madrid, E-28040 Madrid, Spain
\and
Instituto de Astrof\'{\i}sica de Canarias, Avenida V\'{\i}a L\'actea, 38200 La
Laguna, Tenerife, Islas Canarias, Spain
\and
University of Central Florida, Dept. of Physics, PO Box 162385, Orlando,
FL~32816-2385,~USA}
\date{Received 15 July 2008 / Accepted 2 September 2008}

% \abstract{}{}{}{}{} 
% 5 {} token are mandatory
\abstract
{}  
{We investigated in detail nine sources in the direction of the young
$\sigma$~Orionis cluster, which is considered a unique site for studying
stellar and substellar formation.
The nine sources were selected because of some peculiar properties, such as
extremely red infrared colours or too strong H$\alpha$ emission for their
blue optical colours.}   
{We took high-quality, low-resolution spectroscopy (R $\sim$ 500) of the nine
targets with ALFOSC at the Nordic Optical Telescope.
We also re-analyzed [24]-band photometry from MIPS/{Spitzer} and compiled the
best photometry available at the $V i J H K_s$ passbands and the four
IRAC/{Spitzer} channels for constructing accurate spectral energy distributions
covering from 0.55 to 24\,$\mu$m.} 
{The nine targets were classified into: one Herbig Ae/Be star with a scatterer
edge-on disc, two G-type stars, one X-ray flaring, early-M, young star with
chromospheric H$\alpha$ emission, one very low-mass, accreting, young
spectroscopic binary, two young objects at the brown dwarf boundary with the
characteristics of classical T~Tauri stars, and two emission-line galaxies, 
one undergoing star formation, and another one whose spectral energy
distribution is dominated by an active galactic nucleus. 
Besides, we discover three infrared sources associated to overdensities in a
cold cloud in the cluster centre.} 
{Low-resolution spectroscopy and spectral energy distributions are a vital
tool for measuring the physical properties and the evolution of young stars and
candidates in the $\sigma$~Orionis cluster.} 
\maketitle
%
%________________________________________________________________

\section{Introduction}
\label{introduction}

%______________________________________________Fig.
\begin{figure*}
\centering
\caption{False-colour composite images roughly centred on $\sigma$~Ori at
different wavelengths.
From left to right: optical photographic $B_J + R_F + I_N$, 2MASS $J + H +
K_{\rm s}$, and {\em Spitzer} $[3.6] + [8.0] + [24]$ (blue + green + red).
Approximate size is $30 \times 30$\,arcmin$^2$; 
north is up, east is left.
The nine analyzed targets are encircled.
Note the relative dimming of the Trapezium-like star system and the appearence
of a large dust cloud eastwards of it at the reddest wavelengths (see
Sect.~\ref{sigOriIRSs}).
Colour version of all our figures are available in the electronic publication.
{\bf Note: the images are available only in the A\&A version.}}
% ds9
\label{im_all}
\end{figure*}

The \object{$\sigma$~Orionis cluster} is one of the very few star-forming
regions within 1\,kpc with massive stars (B2 and earlier -- Zinnecker \& Yorke
2007). 
In particular, the Trapezium-like system \object{$\sigma$~Ori}, that gives the
name to the cluster, contains five known stars with masses above
$\sim$7\,$M_\odot$ (A, B, D, E, and the recently confirmed F component --
Caballero 2007a and references therein; D.~M. Peterson et~al., in~prep.).
The nearby \object{Horsehead Nebula}, which {\em mane} is illuminated by the
Trapezium-like system, acts as a retaining wall to the east, while approximately
80\,\% of the cluster members are contained in a circular pool of only
20\,arcmin radius, where the extinction is very low (Lee 1968; B\'ejar et~al.
2004; Caballero 2008a).
Besides, although there are still uncertainties on its actual heliocentric
distance ($d \sim$ 350--440\,pc -- Caballero 2008b; Mayne \& Naylor 2008; Sherry
et~al. 2008), $\sigma$~Orionis has been classified as the closest massive-star
forming region (Zinnecker \& Yorke 2007). 
Furthermore, with $\sim$3\,Ma (Zapatero Osorio et~al. 2002a; Oliveira et~al.
2006; Caballero 2008b), the cluster has an age intermediate
between those of extremely young, highly embedded ($\sim$1\,Ma: 
\object{Orion Nebula Cluster}, \object{NGC~2024}) and relatively cleared, still
very young clusters ($\sim$5--10\,Ma: \object{$\lambda$~Orionis},
\object{NGC~2023}) at similar heliocentric distances ($d \sim$~400\,pc).  

The low absorption, relative closeness, and youth make $\sigma$~Orionis to
be one of the most fruitful hunting grounds for substellar objects (B\'ejar
et~al. 1999, 2001, 2004; Scholz \& Eisl\"offel 2004; Kenyon et~al. 2005). 
Notably, the cluster accounts for more than a half of all known {\em confirmed}
isolated planetary-mass objects in the literature (Zapatero Osorio et~al. 2000,
2002b, 2002c, 2007; Barrado y Navascu\'es et~al. 2001; Mart\'{\i}n et~al.
2001a; Gonz\'alez-Garc\'{\i}a et~al. 2006; Caballero et~al. 2007; Scholz \&
Jayawardhana 2008; Luhman et~al. 2008). 
The reader can find in Caballero (2008c) additional cluster data and useful
references for going deeply into all subjects that have been investigated in
$\sigma$~Orionis (from X-ray emission, through Herbig-Haro objects, to brown
dwarfs with~discs). 

For a deeper understanding of the processes that take place in the cluster,
Caballero (2008c) tabulated, in his Mayrit catalogue, 241 $\sigma$~Orionis
stars and brown dwarfs with known feaures of youth (e.g. OB spectral types,
Li~{\sc i} $\lambda$6707.8\,{\AA} in absorption, abnormal strength of other
alkali lines because of low surface gravity, H$\alpha$ $\lambda$6562.8\,{\AA} in
strong emission due to accretion, infrared flux excess due to surrounding disc),
97 candidate cluster members, and 115 back-  and foreground sources.
Of the 338 cluster members and member candidates, 54 are fainter than the
stellar/substellar boundary at $J \approx$ 14.5\,mag for null extinction
(Caballero et~al. 2007).  
The Mayrit catalogue is to date the most comprehensive database of
$\sigma$~Orionis members, and is being used as an input to investigate the
cluster spatial distribution, the frequency of discs at different mass
intervals, the X-ray emission, the  occurrence of wide binaries, or the
initial mass function from about 18\,$M_\odot$ to a few Jupiter masses
(Caballero 2008b; Luhman et~al. 2008; L\'opez-Santiago \& Caballero 2008).
Any systematic error in the catalogue (e.g. incompleteness, contamination) may
have a pernicious influence on the previous results.
It is therefore necessary to confirm or discard membership in the
$\sigma$~Orionis cluster of many Mayrit sources that lack spectroscopy.
Some of these disputable sources have been spectroscopically followed-up in the
very recent works by Maxted et~al. (2008), Gatti et~al. (2008), and Sacco et~al.
(2008).
However, there still remain dozens cluster members and member candidates without
spectroscopy.

\section{Observations and analysis}

\subsection{The target sample}

Here we show low-resolution spectroscopy and spectral energy distributions (from
0.55 to 24\,$\mu$m) of nine selected sources towards the $\sigma$~Orionis
cluster. 
We selected the nine targets based on a variety of properties derived from
existing catalogues (peculiar H$\alpha$ emission, possible extended point
spread function, abnormally red infrared colours, X-ray flares).
The earliest and latest {\em expected} spectral types were B to late M, covering
a magnitude interval of $\Delta V \approx$ 9\,mag (which represents a factor
$\sim$4000 in flux, or a factor $\sim$30 in mass). 
We provide in Table~\ref{observed.targets} the names, coordinates, and
references of the nine targets.  
False-colour composite images centred on each target, with different
wavelength combinations, can be provided upon request to the authors.

\subsection{Low-resolution optical spectroscopy with ALFOSC}

We used the Andaluc\'{\i}a Faint Object Spectrograph and Camera (ALFOSC) at the
2.56\,m Nordic Optical Telescope (NOT) in the Spanish Observatorio del
Roque de  Los  Muchachos during five nights in Oct~2006. 
The ALFOSC detector was the CCD\#8, a nimo back illuminated
E2V~2k$\times$2k~42--40 with a plate scale of 0.19\,arcsec/pixel (2$\times$2
binning).
With the grism~\#5 and the 1.0\,arcsec-size slit, we got a dispersion of
3.1\,\AA/pixel and a resolution of 14.2\,{\AA} at 7000\,{\AA} (R $\equiv \lambda  
/ \Delta \lambda$ = 494).

The total wavelength coverage was 5350--9910\,{\AA}.
The central wavelength (7630\,\AA) was red-shifted with respect to the
effective blaze wavelength (6500\,\AA) to maximise the use of the reddest
part of the optical spectrum, where late M stars (our faintest objects) emit
most of their energy. 
With the grism~\#5, the second order (corresponding to bandpasses $B$ and $V$)
is noticeable at the 20\,\% level only at wavelengths $\lambda >$ 9700\,{\AA}
(no blocking filter was therefore used).
Unfortunately, the peak-to-peak fringe levels start at 6700\,{\AA} and increase in
intensity redwards (8\,\% at 7500\,\AA, 18\,\% at 8000\,\AA, $>$30\,\% at
$>$9000\,\AA).
The fringing, together with the low number of counts of flux standard stars at
the extremes of the total wavelength coverage due to the lower quantum
efficiency of the detector, made the final useful wavelength coverage to
shrink to 5500--9300\,\AA.

The weather during our run was in general fair, with high thin cirruses only on
the last night. 
The seeing ranged in the interval $\sim$0.7--1.0\,arcsec, which justified our
slit size choice.
Wind speed was low and humidity maintained roughly stable within each night (with
maximum variations of 15\,\%).
The dates of observation and the exposure times of the nine (combined) spectra
are provided in  Table~\ref{observed.targets}.
Total exposure times ranged from only 30\,s for the brightest target to
1\,h\,20\,min for the faintest cluster member.
Bias images for posterior calibration were taken at sunsets, while flats and
arcs were obtained just before and after each scientific spectrum.
The spectra of the five faintest targets are actually the combination of two to
four individual spectra, sometimes taken on different nights (e.g. the final
spectrum of Mayrit~264077 is the combination of two 900\,s-long spectra taken on
2006~Oct 11 and 12).
We also got the spectra of \object{GJ~3517} (M9V) during nighttime.
It was used as a flux, spectral type, and radial velocity
standard/comparison~star. 

Reduction of the raw images (bias subtraction and flat-fielding), spectrum
extraction, flux calibration, and combination of the spectra (in this order),
was carried out using standard procedures within the IRAF environment ({\tt
daophot.onedspec}).
We applied a tiny Doppler shift of 2\,{\AA} to the spectra taken during the first
two nights due to a systematic error in the calibration.
Since the resolution power was 14.2\,{\AA} (measured from the arcs and sky
lines), this shift had no implications on further analysis steps.
Before combination, we checked that there were no significant differences
between individual spectra (e.g. no appreciable variation of the H$\alpha$
emission). 

\begin{figure}
\centering
\includegraphics[width=0.53\textwidth]{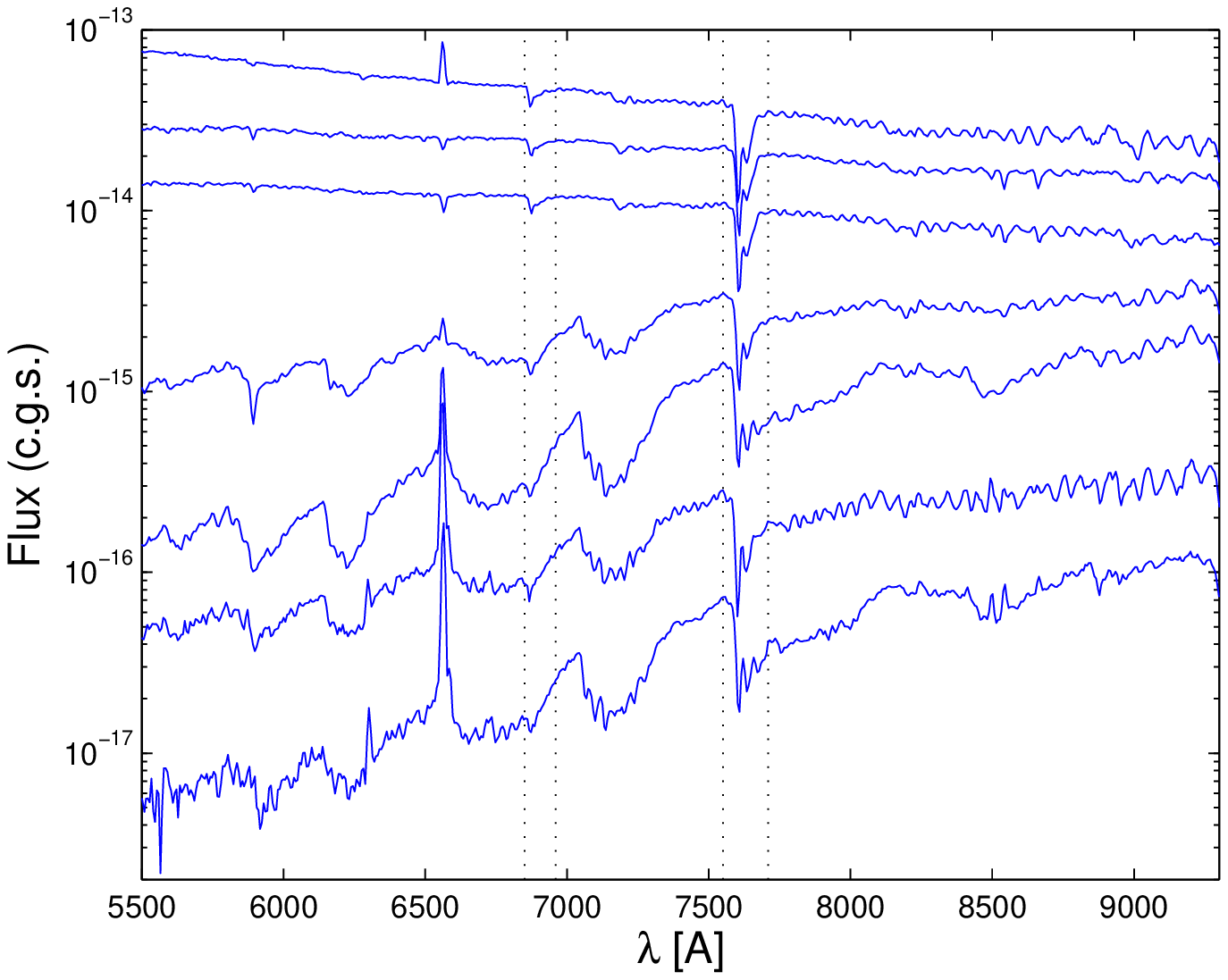}
\caption{Flux-calibrated, low-resolution ALFOSC/NOT spectra of the stars and
brown dwarfs Mayrit~459340, Mayrit~968292, Mayrit~1064167, Mayrit~797272,
Mayrit~102101AB, Mayrit~264077, and Mayrit~358154, from top to bottom.
Vertical dotted lines indicate strong telluric absorption features.
Beyond $\sim$8000\,{\AA}, ripples show regions where fringing contributes to the
spectra.}   
% pintaespectros_new.m
\label{spAtoG}
\end{figure}

\begin{figure}
\centering
\includegraphics[width=0.53\textwidth]{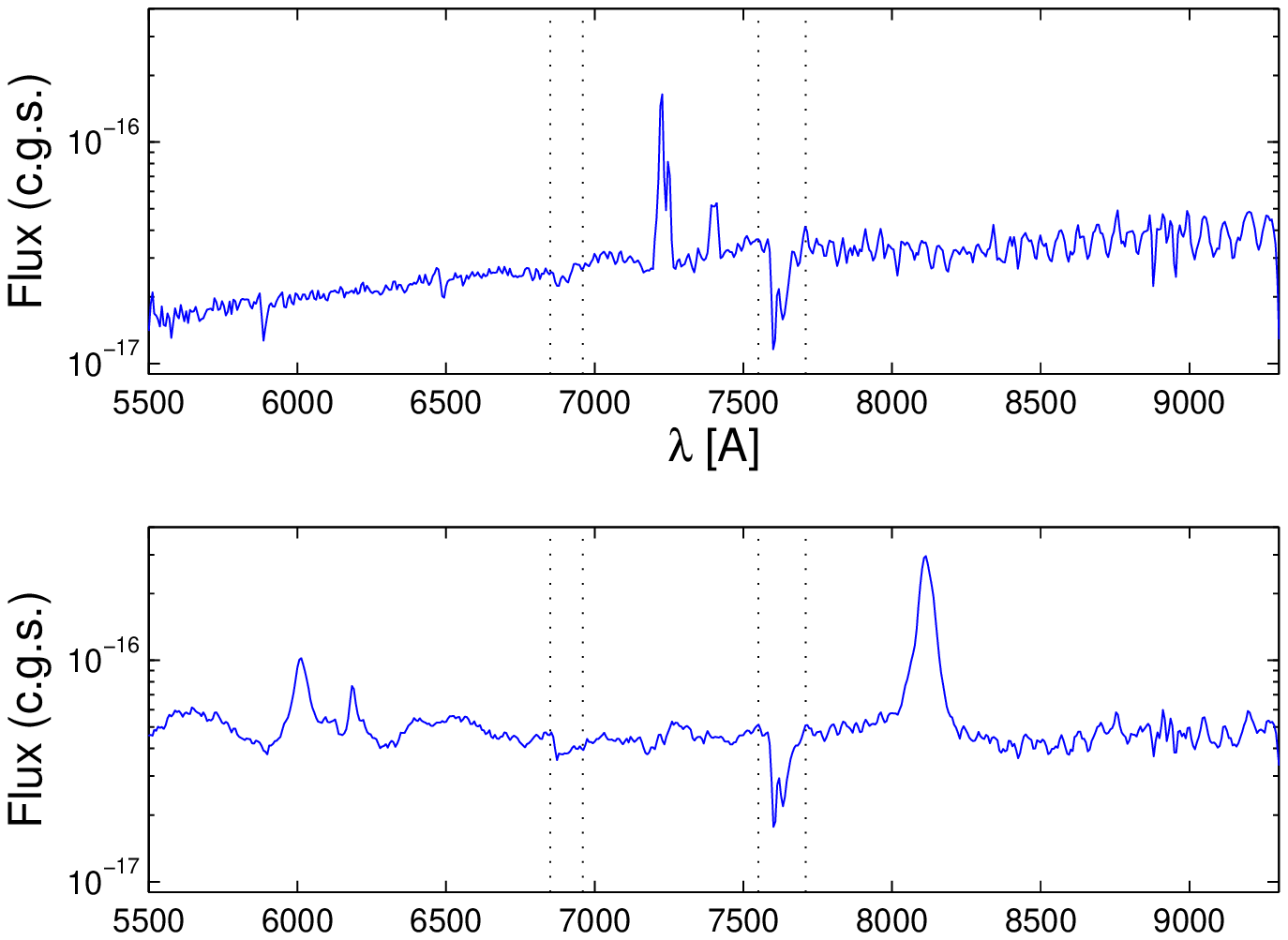}
\caption{Same as Fig.~\ref{spAtoG}, but for the emission-line galaxies 
UCM~0537--0227 (top window) and UCM~0536--0239 (bottom window).}  
% pintaespectros_new.m
\label{spHandI}
\end{figure}

The nine final (combined) spectra are shown in Figs.~\ref{spAtoG}
and~\ref{spHandI}. 
Note the strong telluric absorption features, especially the O$_{2}$  A and
B bands at 6850--6960 and 7550--7710\,{\AA} and another patent water vapour
absorption feature at $\lambda$ = 7150--7340\,{\AA} (there is other strong
feature at $>$9360\,{\AA}, not visible with our display). 

%______________________________________________Fig.
%

%______________________________________________Fig.

\subsection{Multi-wavelenght photometry}

\subsubsection{$V i J H K_{\rm s} [3.6] [4.5] [5.8] [8.0]$-band photometry}

We collected optical, red optical, near-infrared, and mid-infrared
magnitudes from the literature for the nine sources.
The $i$ and $JHK_{\rm s}$ magnitudes were homogeneously retrieved from the
Deep Near Infrared Survey of the Southern Sky (DENIS; Epchtein et~al. 1997) and
the Two Micron All Sky Survey (2MASS; Skrutskie et~al. 2006).
The magnitudes in the four channels of the Infrared Array Camera (IRAC;
3.8--8.0\,$\mu$m) on the {\em Spitzer Space Telescope} were taken either from
Caballero et~al. (2008) or from Hern\'andez et~al. (2007), depending on
availability.
Finally, we compiled the most accurate available visual magnitudes for each
target from a heterogeneous collection of sources.
We give in Table~\ref{photometry.targets} all the necessary information,
including the magnitudes and their origin.

\subsubsection{$[24]$-band photometry with MIPS/{\em Spitzer} and spectral
energy distributions}

%______________________________________________Fig.
\begin{figure*}
\centering
\includegraphics[width=1.10\textwidth]{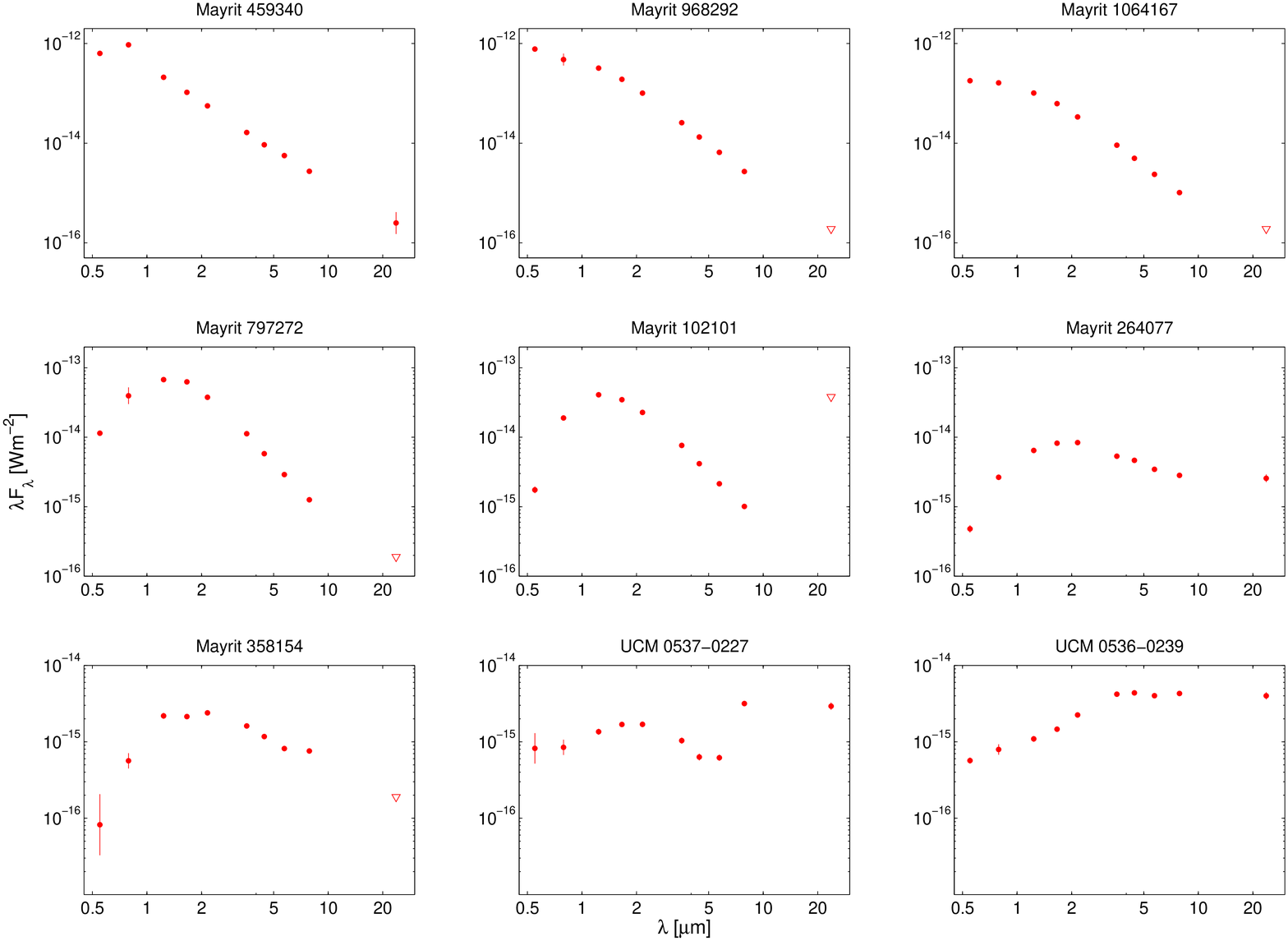}
\caption{Spectral energy distributions of the nine targets in
Tables~\ref{observed.targets} and~\ref{photometry.targets}.  
From the shortest to the longest wavelengths, the data points correspond to the
passbands Johnson $V$, DENIS $i$, 2MASS $J$, $H$, and $K_{\rm s}$, IRAC/{\em
Spitzer} $[3.6]$, $[4.5]$, $[5.8]$, and $[8.0]$, and MIPS/{\em Spitzer} $[24]$.
The lower triangle indicates an upper limit for the flux at the $[24]$
passband.} 
% sed_new.m
\label{sedAtoI}
\end{figure*}

Hern\'andez et~al. (2007) tabulated 24\,$\mu$m Multiband Imaging Photometer for
{\em Spitzer} (MIPS) magnitudes ([24]) for only a few of our targets.
To expand the wavelength coverage of our spectral energy distributions, we
downloaded with the Leopard tool the most recent post-Basic Calibrated Data
$[24]$-band MIPS mosaic covering the whole $\sigma$~Orionis cluster (P.I.:
G.~Rieke, programme: Evolution and Lifetimes of Protoplanetary Disks).

First, we looked for sources suitable for studying their observed $[24]$-band
flux. 
As can be seen in Caballero et~al. (2008) and Hern\'andez et~al. (2007),
intracluster clouds (some of them associated to the Horsehead Nebula), hardly
visible in the optical and in the near-infrared, are apparent at wavelengths
redwards of $\sim$8\,$\mu$m (see Sect.~\ref{sigOriIRSs}). 
To avoid contamination by spurious detections, we imposed a relatively high
threshold for the source finding at 10\,$\sigma$ over the noise background.
Next, we performed standard aperture and point-spread-function (PSF) photometry
on the mosaic with the IRAF package {\tt digiphot}, using an aperture radius of
13\,arcsec and a background annulus from 39 to 48\,arcsec (which leaded to an
aperture correction of 1.175 according to the {\em Spitzer} Science
Centre\footnote{\tt http://ssc.spitzer.caltech.edu/mips/apercorr/.}).
We used $F_{\nu0}$ = 7.17$\pm$0.11\,Jy and $\lambda_0$ = 23.675\,$\mu$m
for transforming PSF fluxes into apparent magnitudes.
In the previous-to-the-last column in Table~\ref{photometry.targets}, we give
the measured [24] magnitudes and their errors, computed from the quadratic sum
of the error in the PSF photometry and the errors in the aperture correction
and the absolute calibration ($\sim$7\,\%; Engelbracht et~al. 2007).
The 10\,$\sigma$ threshold made the completeness limit of our $[24]$-band survey
to be $\sim$9.2\,mag.

Only the magnitudes for five of our nine main targets were obtained.
Besides, the tabulated magnitude for one of them (Mayrit~102101AB) actually
corresponds to a bright knot in a cloud in the north-east vicinity of
$\sigma$~Ori, spatially-coincident with our star (see again
Sect.~\ref{sigOriIRSs}). 
The other four stars without MIPS measurement have, therefore, magnitudes $[24]
\gtrsim$ 9.2\,mag.

Using the magnitudes in Table~\ref{photometry.targets} and the zero-point
values for conversion between magnitudes and flux densities compiled in the
NASA/IPAC/MSC Star and Exoplanet Database\footnote{\tt 
http://nsted.ipac.caltech.edu/NStED/docs/parhelp/Photometry.html.}, we
constructed the spectral energy distributions shown in Fig.~\ref{sedAtoI}.
%For Mayrit~459340, we plot Tycho-2 $V_T$ instead of Johnson $V$ (accounting for
%the different central wavelengths [$\lambda_0$] and conversion factors between
%flux densities and magnitudes [$F_{\nu 0}$]).
%The bluest data point corresponds to Tycho-2 $B_T$.    

\section{Results and discussion}

We have classified our nine targets into three categories, depending on their
astrophysical nature.
The classes are early-type and solar-like stars (3), late-type young stars
and brown dwarfs (4), and emission-line galaxies at intermediate redshifts (2).
In the last columns of Table~\ref{observed.targets}, we list the spectral
types, equivalent widths of H$\alpha$, and cluster membership measured by us of
the seven young object candidates.
Actually, we measured the pseudo-equivalent width (pEW) of the lines with
respect to the pseudo-continuum of the four M-type stars and brown dwarfs.
This definition extends to other lines different from the Balmer series (e.g.
Li~{\sc i}).
Hereafter, we use the term equivalent width (EW) for simplicity.
The error in the EWs comes from the uncertainty in the determination of the
(pseudo-)continuum.

The seven young object candidates and the two galaxies are discussed in detail
in the following sections.

\begin{sidewaystable*}
\begin{minipage}[t][180mm]{\textwidth}
      \caption[]{Names, coordinates, and basic spectroscopic data of 
      observed targets.} 
         \label{observed.targets}
\centering
         \begin{tabular}{ll cc l lc lc l}
            \hline
            \hline
            \noalign{\smallskip}
Name$^{a}$			& Alternative			& $\alpha$ 	& $\delta$      & Reference(s)$^{b}$	 & Date 	 & $t_{\rm exp}$	& Sp.  	 		& EW(H$\alpha$)	& Cluster \\ % & Remarks$^{a}$ 		       
				& name				& (J2000) 	& (J2000)       &		       	 &		 & (s)  		& type 	 		& [\AA]		& member  \\ % &			       
            \noalign{\smallskip}															   					 	   		      
            \hline
            \noalign{\smallskip}															   					 	   		      
\object{Mayrit~459340}		& StHA~50			& 05 38 34.45 	& --02 28 47.6  & St86, DK88, Ca07a      & 2006 Oct 15   & 30			& A2--6 V e		& --10$\pm$1	& Yes     \\ % & H$\alpha$, Be?	       
\object{Mayrit~968292}		& GSC~04771--00962		& 05 37 44.92	& --02 29 57.3  & Ca06, Ca07a 	       	 & 2006 Oct 14   & 60			& G8--K0 V		& +2.5$\pm$0.5 	& Yes?    \\ % & X?, wide binary?		       
	Mayrit~1064167		& \object{2MASS~J05390088--0253164}& 05 39 00.88& --02 53 16.4  & ...		       	 & 2006 Oct 12   & 150  		& G2--5 V		& +3.5$\pm$0.5 	& No?     \\ % & New				       
\object{Mayrit~797272}		& [W96]~rJ053751--0235		& 05 37 51.61 	& --02 35 25.7  & Sh04, Fr06, Ca06       & 2006 Oct 13   & 300  		& M1--3 V e		&--4.5$\pm$0.5 	& Yes     \\ % & XX, M0--2, Li {\sc i}, H$\alpha$?	 
\object{Mayrit~102101}AB	& [W96]~rJ053851--0236		& 05 38 51.45 	& --02 36 20.6  & Wo96, Fr06, He07, Sa08 & 2006 Oct 11, 13& 300+500		& M2--4 V e		& --32$\pm$2	& Yes     \\ % & H$\alpha$, X, SB2, ev.~disc	       
\object{Mayrit~264077}		& S\,Ori~J053902.1--023501	& 05 39 01.94 	& --02 35 02.9  & He07, Ca07	       	 & 2006 Oct 11, 12& 900+900		& M2--4 V e		& --72$\pm$4	& Yes     \\ % & mIR, variable?	       
\object{Mayrit~358154}		& S\,Ori~J053855.4--024121	& 05 38 55.42 	& --02 41 20.8  & He07, Ca07	       	 & 2006 Oct 13, 15& 1200+3$\times$1200  & M4--6 V e 		&--190$\pm$20  	& Yes     \\ % & mIR				       
\object{UCM~0537--0227}		& Mayrit~926051			& 05 39 32.70 	& --02 26 15.4  & Ca08  	       	 & 2006 Oct 11, 12& 1200+1200		& ...$^{c}$		& ...$^{c}$	& No	   \\ % & extended?			       
\object{UCM~0536--0239}		& [HHM2007] 668			& 05 38 41.24 	& --02 37 37.7  & He07, Ca07b	       	 & 2006 Oct 11, 12& 1200+1200		& ...$^{c}$		& ...$^{c}$	& No	   \\ % & X, class~I?, extended?	       
            \noalign{\smallskip}
            \hline
         \end{tabular}
\begin{list}{}{}
\item[$^{a}$] Mayrit and UCM designations follow the nomenclatures
introduced in Caballero (2007b) and Zamorano et~al. (1994), respectively
(truncated coordinates in the UCM~HHMM+DDMMW designation are for the 1950.0
equinox).
\item[$^{b}$] References --
St86: Stephenson (1986);
DK88: Downes \& Keyes (1988);
Sh04: Sherry, Walter \& Wolk (2004);
Fr06: Franciosini, Pallavicini \& Sanz-Forcada (2006);
He07: Hern\'andez et~al. (2007);
Ca07a: Caballero (2007a);
Ca07b: Caballero (2007b).
Ca07: Caballero et~al. (2007);
Ca08: Caballero (2008c);
Sa08: Sacco et~al. (2008).
\item[$^{c}$] Not applicable for galaxies UCM~0537--0227 and 
UCM~0536--0239.
\end{list}
\vfill
\end{minipage}
\end{sidewaystable*}

\begin{sidewaystable*}
\begin{minipage}[t][180mm]{\textwidth}
      \caption[]{Photometry of observed targets.} 
         \label{photometry.targets}
\centering
         \begin{tabular}{l c c ccc cccc c c}
            \hline
            \hline
            \noalign{\smallskip}
Name			& $V \pm \delta V$ 	& $i \pm \delta i$ 	& $J \pm \delta J$      & $H \pm \delta H$      & $K_{\rm s} \pm \delta K_{\rm s}$ & [3.6]	& [4.5]			& [5.8]			& [8.0]			& [24.0]		& Sources of \\  
			& [mag]			& [mag]			& [mag]		        & [mag] 	        & [mag] 		& [mag] 		& [mag]			& [mag]			& [mag]			& [mag]			& photometry$^{a}$ \\  
            \noalign{\smallskip}
            \hline
            \noalign{\smallskip}
Mayrit~459340		& 11.28$\pm$0.05      	& 11.004$\pm$0.040      & 10.666$\pm$0.024      & 10.614$\pm$0.024      & 10.545$\pm$0.024	& 10.40$\pm$0.03	& 10.29$\pm$0.03	& 10.07$\pm$0.03	&  9.88$\pm$0.04	& 8.90$\pm$0.55		& A122244440	\\ % Ca inp., V from Tycho-2 
Mayrit~968292		& 11.06$\pm$0.07      	& 10.747$\pm$1.000      & 10.202$\pm$0.026      &  9.960$\pm$0.022      &  9.905$\pm$0.022	&  9.91$\pm$0.03	&  9.90$\pm$0.03	&  9.91$\pm$0.03	&  9.89$\pm$0.03	& $\gtrsim$9.2$^b$	& A122233330	\\ % He07, V from Tycho-2     
Mayrit~1064167		& 12.65$\pm$0.12	& 11.912$\pm$0.030      & 11.451$\pm$0.024      & 11.184$\pm$0.023      & 11.101$\pm$0.024	& 11.04$\pm$0.04	& 10.97$\pm$0.04	& 11.01$\pm$0.05	& 10.95$\pm$0.07	& $\gtrsim$9.2		& B122244440	\\ % Ca inp., V from ASAS-3 
Mayrit~797272		& 15.64$\pm$0.01      	& 13.446$\pm$1.000      & 11.894$\pm$0.026      & 11.172$\pm$0.023      & 10.977$\pm$0.022	& 10.81$\pm$0.03      	& 10.80$\pm$0.03      	& 10.79$\pm$0.03      	& 10.72$\pm$0.03      	& $\gtrsim$9.2		& 3122233330	\\ % He07    
Mayrit~102101AB		& 17.68$\pm$0.09      	& 14.248$\pm$0.030      & 12.438$\pm$0.027      & 11.813$\pm$0.023      & 11.552$\pm$0.025	& 11.23$\pm$0.03      	& 11.16$\pm$0.03      	& 11.12$\pm$0.03      	& 10.96$\pm$0.03      	& 3.44$\pm$0.27$^c$	& 3122233330	\\ % He07    
Mayrit~264077		& 19.08$\pm$0.13      	& 16.380$\pm$0.080      & 14.445$\pm$0.035      & 13.378$\pm$0.030      & 12.607$\pm$0.030	& 11.45$\pm$0.04	& 10.99$\pm$0.04	& 10.52$\pm$0.04	&  9.75$\pm$0.04	& 6.37$\pm$0.14		& C122255550	\\ % Ca07, V from Wolk (1996), He07: 11.62$\pm$0.03	& 11.04$\pm$0.03	& 10.60$\pm$0.03	&  9.84$\pm$0.03 
Mayrit~358154		& $\gtrsim$20.0		& 18.063$\pm$0.250      & 15.622$\pm$0.096      & 14.842$\pm$0.050      & 13.968$\pm$0.061	& 12.88$\pm$0.08	& 12.52$\pm$0.08	& 12.08$\pm$0.09	& 11.15$\pm$0.08	& $\gtrsim$9.2$^d$	& D122255550	\\ % Ca07; He07: & 12.88$\pm$0.08	& 12.52$\pm$0.08	& 12.08$\pm$0.09	& 11.15$\pm$0.08	 & 12.92$\pm$0.03	& 12.54$\pm$0.03	& 12.17$\pm$0.03	& 11.27$\pm$0.03
UCM~0537--0227		& $\sim$18.5		& 17.622$\pm$0.250      & 16.141$\pm$0.098      & 15.099$\pm$0.071      & 14.345$\pm$0.092	& 13.40$\pm$0.10	& 13.21$\pm$0.11	& 12.47$\pm$0.10	&  9.72$\pm$0.04	& 6.23$\pm$0.12		& D122244440	\\ % Ca inp. 
UCM~0536--0239		& 18.90$\pm$0.10	& 17.693$\pm$0.170      & 16.375$\pm$0.094      & 15.253$\pm$0.080      & 14.040$\pm$0.058	& 11.88$\pm$0.03	& 11.11$\pm$0.03	& 10.44$\pm$0.03	&  9.39$\pm$0.03	& 5.89$\pm$0.12		& E122233330	\\ % He07, V from JOSE 
            \noalign{\smallskip}
            \hline
         \end{tabular}
\begin{list}{}{}
\item[$^{a}$] Sources of photometry --
0: this work;
1: DENIS;
2: 2MASS;
3: Hern\'andez et~al. (2007);
4: Caballero et~al. (2008); % poster Pasadena
5: Caballero et~al. (2007); % A&A
A: transformed from $V_T$ Tycho-2 (H{\o}g et~al. 2000) to $V$ Johnson using 
relations in ESA (1997);
B: ASAS-3 (Pojma\'nski 2002);
C: Wolk (1996);
D: estimated from photographic $B_J$ and $R_F$ magnitudes;
E: measured using unpublished $V$ data obtained with the Wide Field Camera at 
the Isaac Newton Telescope, calibrated into the Johnson system through  
Tycho-2 comparison stars in an overlapping short exposure $V$ image in 
Caballero~(2007b).
\item[$^{b}$] Hern\'andez et~al. (2007) measured $[24]$ = 9.80$\pm$0.05\,mag for Mayrit~968292;
it is fainter than the 10\,$\sigma$ threshold in our data (note their under
estimated errors). 
\item[$^{c}$] The $[24]$ emission of Mayrit~102101AB actually comes from a
spatially-coincident intracluster cloud described in the text. 
\item[$^{d}$] Hern\'andez et~al. (2007) measured $[24]$ = 8.58$\pm$0.03\,mag for
Mayrit~264077; 
this emission actually comes from a nearby source (which magnitude we
re-determine at $[24]$ = 8.32$\pm$0.37\,mag). 
\end{list}
\vfill
\end{minipage}
\end{sidewaystable*}

\subsection{Early-type and solar-like stars}

\subsubsection{Mayrit~459340} 
\label{M459340}

Mayrit~459340 (StHA~50) was a new H$\alpha$ emission star found above 10\,deg
galactic latitude in Stephenson (1986), who investigated a large number of
red-sensitive objective prism plates.  
Soon afterwards, Downes \& Keyes (1988) classified Mayrit~459340 as a Be star
from a spectrum taken with an image tube scanner.
It went unnoticed until Caballero (2007a) listed it as one of the (30) brightest
stars of the $\sigma$~Orionis cluster (Mayrit~459340, with $M \gtrsim$
1.4\,$M_\odot$, is also one the most massive cluster stars).
According to him, Mayrit~459340 deviates at the level of 3\,$\sigma$ to the blue
of the spectro-photometric cluster sequence in a $V$ vs. $B-V$ colour-magnitude
diagram from Tycho-2 magnitudes. 
Although its early (Be) spectral type matches, within uncertainties, its blue
colours in the optical ($B_T - V_T$ = --0.06$\pm$0.11\,mag), % $B_T - V_T$ = --0.056$\pm$0.107\,mag
Mayrit~459340 is $\Delta B_T$ = 1.96$\pm$0.06\,mag % $\Delta B_T$ = 1.956$\pm$0.062\,mag
fainter than the faintest B-type star in $\sigma$~Orionis (HD~294275
[\object{Mayrit~1227243}, B9V]; Caballero 2007a).
Its $V_T$ magnitude better matches that of an early- or mid-F
spectral-type star in the cluster, or a late-B spectral-type star at an unlikely
heliocentric distance thrice larger than to $\sigma$~Orionis.
The first hypothesis would easily explain its H$\alpha$ emission, while the
second hypothesis would locate Mayrit~459340 at about 300\,pc below the Galactic
plane (the Galactic vertical scale height of OBA-type stars is only about
100\,pc; e.g. Bahcall \& Soneira~1980). 

Likewise, Caballero (2007a) also reported that Mayrit~459340 has a disc based on
3.6--8.0\,$\mu$m IRAC/{\em Spitzer} photometry from Caballero et~al. (2008).
Our MIPS/{\em Spitzer} $[24]$ photometry (tabulated in
Table~\ref{photometry.targets}) corroborated his statement: 
we derived colours $[3.6]-[24] = 1.5 \pm 0.6$\,mag and $[3.6]-[8.0] = 0.52 \pm
0.05$\,mag, that deviates about 3\,$\sigma$ and 10\,$\sigma$ with respect the
expected values for a disc-less dwarf  ($[3.6]-[24] \sim [3.6]-[8.0] \sim
0$\,mag). 
From the ALFOSC spectrum, shown in the top of Fig.~\ref{spAtoG}, we
measured a very intense H$\alpha$ emission of --10$\pm$1\,{\AA} and derived a
spectral type in the interval A2--6V (that corresponds to a most probable mass
in the interval 3--4\,$M_\odot$).  
This spectral type approximately matches the blue $B_T - V_T$ colour of
Mayrit~459340 and the previous determination by Downes \& Keyes (1988).  
For the classification, we investigated the depth of the blended doublets
He~{\sc i}~D$_3$ $\lambda\lambda$5875.6,5876.0\,{\AA} and Na~{\sc i}
$\lambda\lambda$5890.0,5895.6\,{\AA} and of the line Ca~{\sc i}
$\lambda$5857.5\,{\AA} and compared the slope of the continuum of its spectrum
with those of early-type standard stars from Valdes et~al. (2004).
The strong H$\alpha$ emission seems to persist in time, since 1978--1984
(Stephenson 1986), through 1986--1987 (Downes \& Keyes 1988), to the present
(this work).
Besides, the absorption feature of the diffuse interstellar band at about
6280\,{\AA}, typical in early-type stars in young clusters, is quite clear.
To our knowledge, there is only one scenario that simultaneously explain the A
spectral type, the H$\alpha$ emission, the abnormal blue colour for its
magnitude (or, alternatively, its abnormal faintness for its colour), and the
flux excess at 8--24\,$\mu$m: the {\em blueing} effect in a Herbig Ae/Be (HAeBe)
star with a scatterer edge-on disc (Th\'e et~al. 1994 and references
therein).
  
There are grounds to consider Mayrit~459340 a (long-scale) photometric
variable, which is another characteristic typical of HAeBe stars. 
From data of the All Sky Automated Survey (ASAS-3; Pojma\'nski 2002),
we measured $\overline{V} = 11.312$\,mag, $\sigma_V = 0.045$\,mag, and
$\overline{\delta V} = 0.032$\,mag ($N$ = 248), consistent with the Tycho-2
$V_T$ measurement and with Mayrit~459340 lacking large-amplitude photometric
variations ($\Delta V \gg$ 0.045\,mag) during years 2000 to 2005.
There is not evidence of photometric variability in the USNO-B1.0 catalogue,
either (Monet et~al. 2003).
However, from the spectral energy distribution, the DENIS $i$ magnitude,
measured in 1996 Feb, seems to be $\sim$0.6--0.7\,magnitudes brighter than
expected. 
The value $i = 11.00 \pm 0.04$\,mag is still far away from the $i \sim
10.0$\,mag threshold below which brighter DENIS stars in $\sigma$~Orionis are
affected by saturation and non-linear effects (Caballero 2008c).
From the ratio between our flux-calibrated spectra, we also estimated that
during the observations (2006 Oct), Mayrit~459340 was $\sim$0.9\,mag {\em
brighter} at 6200--6300\,{\AA} than the solar-like, young star candidate
Mayrit~968292. 
However, from the compiled magnitudes in Table~\ref{photometry.targets},
Mayrit~459340 is only brighter than Mayrit~968292 at $\lambda \gtrsim$
8\,$\mu$m (where the flux excess by the disc gets more important).
Furthermore, the tabulated $V_T$ magnitudes are contemporaries (early 1990s),
and Mayrit~459340 was 0.17$\pm$0.12\,mag {\em fainter} than Mayrit~968292 at the
corresponding effective wavelength of 5320\,\AA. 
% $\overline{\lambda}(V_T)$ = 5319\,\AA
This apparent brightening at recent epochs, if accompanied by a suitable colour
change, would make Mayrit~459340 to lie on the right spectro-photometric
sequence of $\sigma$~Orionis.
Recalling the scatterer edge-on disc scenario of Th\'e et~al. (1994),
``the variations in brightness and in colour [of a HAeBe star] are caused by a
dust cloud in a Keplerian orbit revolving in the outer part of a star's
circumstellar disc''. 
Most of the time, the disc of Mayrit~459340 is oriented almost edge on
with respect to the observer, and the star is well obscured by the dust cloud.
It is at this stage when the dimming is maximum, and scattered blue light is
more easily detected.
It is expected that when the dust cloud passes away from the front of the HAeBe
star, it gets brighter and redder.
This effect has been clearly detected in the A3e-type, variable star (with rapid
variations) UX~Ori (Bibo \& Th\'e 1991).
Mayrit~459340 is about 2.6\,mag fainter than UX~Ori, which probably had
made it to get unnoticed until now.
High-resolution spectroscopy would enable better constraints on membership
through measurments of radial velocity.

\subsubsection{Mayrit~968292 and Mayrit~1064167} 

On the one hand, the Tycho-2 star Mayrit~968292 (GSC~04771--00962) was the
brightest (in the optical) $\sigma$~Orionis member candidate in the Mayrit
catalogue {\em without} spectroscopy (Caballero 2008c).
This was a {\em per~se} reason for first obtaining an optical spectrum with
ALFOSC.
With $V \approx$ 11.1\,mag, it was necessary to exposure only during 1\,min.
Besides, Mayrit~968292 is the brightest component of a visual binary with a
relatively small angular separation ($\rho \approx$ 9.8\,arcmin, $\theta
\approx$ 97\,deg, $\Delta J = 0.72 \pm 0.03$\,mag).
The secondary, \object{Mayrit~958292}, displayed an apparent lithium absorption
feature in a spectrum taken by Caballero (2006), which confirmed the star to be
a very young member of the Ori~OB1b association.
If both stars formed a physical pair, then it would be one of the very few wide
binaries ($r \approx$ 3800\,AU) detected so far in the $\sigma$~Orionis cluster
(several wide binary candidates were listed in Caballero [2007b]).
See also Caballero (2007a) for a consideration on the possible X-ray emission of
Mayrit~968292.

On the other hand, Mayrit~1064167 (2MASS~J05390088--0253164) was an anonymous
star whose position in the $i$ vs. $i-K_{\rm s}$ diagram made it to fall in the
side of the candidate fore- and background stars, but still very close to the
sequence of the cluster (see fig.~4 in Caballero~2008c).
It is the brightest one in a trio of stars separated by 20--40\,arcsec.
One of the other two components of the asterism has also optical colours and
magnitudes typical of solar-like cluster members (e.g. Sherry et~al. 2004).
However, Caballero (2006) found no trace of lithium in a high signal-to-noise,
mid-resolution spectrum of the target.
The remaining star in the trio is a foreground late-type dwarf or giant based on
a relatively large proper motion and quite red optical/near-infrared colours for
its magnitude ($J-K_{\rm s} = 1.30 \pm 0.04$\,mag; Caballero 2008c). 
Far for extrapolating the non-membership also to Mayrit~1064167, we found three
X-ray events tabulated by the Second {\em ROSAT} Position-Sensitive Proportional
Counter (PSPC) Catalogue, 2RXP (ROSAT 2000), and the White-Giommi-Angelini
version of the {\em ROSAT} PSPC Catalogue, WGACAT (White et~al. 2000), at
10--30\,arcsec to the star.
Although the X-ray events could also be associated to a background extragalactic
source, we decided to test photometric selection criteria for cluster members in
the literature (in particular: Sherry et~al. 2004; Caballero~2008c) with the
possible X-ray emitter Mayrit~1064167. 

At a first glance at the ALFOSC spectra and the spectral energy
distributions of Mayrit~968292 and Mayrit~1064167, the two stars seemed to be
G-type stars with no indication of accretion or infrared flux excess.
We could only impose lower limits to the $[24]$-band magnitudes.
A careful comparison of the blended Na~{\sc i}~D$_1$D$_2$ doublets and
H$\alpha$ lines indicated that Mayrit~968292 is cooler than Mayrit~1064167 (the
absorption of the Na~{\sc i} doublets and the H$\alpha$ line increases and
decreases, respectively, towards later spectral types).   
By comparison with several spectral-type standard stars taken from Jacoby et~al.
(1984), Montes et~al. (1997), and Valdes et~al. (2004), we classified
Mayrit~968292 as G8--K0V and Mayrit~1064167 as G2--G5V. 
We obtained a similar classification using the lines ratio H$\alpha$/Fe~{\sc i}
$\lambda$6495.0\,{\AA} (Danks \& Dennefeld 1994).
Other lines typical of solar-like stars that we observed in the spectra of
Mayrit~968292 and Mayrit~1064167 were the calcium features Ca~{\sc i}
$\lambda$6163.8\,{\AA} and, especially, the infrared triplet Ca~{\sc ii}
$\lambda\lambda\lambda$8498.0,8542.1,8662.1\,{\AA} (partially embedded in the
fringing region).
Unfortunately, our spectra have not enough spectral resolution to clearly detect
any other spectroscopic feature of youth (especially lithium).

Although Mayrit~1064167 has an earlier spectral type than Mayrit~968292, the
first star is about 1.2\,mag fainter than the second one at all passbands.
Since such a large reversal of brightness is improbable, we
consider that at least one of the two stars is actually a contaminant.
The interloper is probably Mayrit~1064167, because Mayrit~968292 follows the
cluster sequence and Mayrit~1064167 does not (it has roughly the same magnitudes
of mid-K-type young stars in $\sigma$~Orionis). 
Albeit Mayrit~968292 remains as a good solar-like cluster member candidate, it
is necessary to obtain higher-resolution spectroscopy to confirm its membership. 
If its youth is verified, its membership in the Mayrit~968292--958292 binary
will even so remain unknown for a longer time: because of of the location of
$\sigma$~Orionis in the solar antapex, its cluster members have very low proper
motions ($\mu <$ 10\,mas\,a$^{-1}$; Caballero 2007a)

\subsection{Late-type young stars and brown dwarfs}

We mostly used the PC3 index (Mart\'{\i}n, Rebolo \& Zapatero Osorio 1996;
Mart\'{\i}n et~al. 1999) for the spectral type classification of the
four M-type objects, although we also took into account the information
provided by the I1, I2, and I3 indices (indicators of the strength of several
CaH and TiO absorption bands; Mart\'{\i}n \& Kun 1996) and some indices
associated to metallic hydrides and oxides (CrH, FeH, H$_2$O, TiO, and VO). 
We also compared the indices values with those of spectral standards in
the literature and of GJ~3517, observed with the same instrumental
configuration.

\subsubsection{Mayrit~797272} 

Mayrit~797272 ([W96]~rJ053751--0235) is a moderate X-ray emitter, early-M
star in $\sigma$~Orionis. % [SWW2004]~125, [FPS2006]~1
It was first discovered by Wolk (1996), who classified it as a {\em ROSAT} M0
star with lithium in absorption (EW(Li~{\sc i}) = +0.32\,\AA) and H$\alpha$ in
faint chromospheric emission (EW(H$\alpha$) = --6.4\,\AA).
A decade later, Franciosini et~al. (2006) measured its X-ray count rate with
{\em XMM-Newton} and estimated an M2 spectral type from the $R-I$, $V-I$, and
$I-J$ colours using the relations by Kenyon \& Hartmann (1995) and Leggett
et~al. (2001).  
Almost simultaneously, Caballero (2006) took a low signal-to-noise, optical
spectrum where the H$\alpha$ line appeared in faint, broadened emission and the
Li~{\sc i} line was undetected.
Very recently, L\'opez-Santiago \& Caballero (2008) have found Mayrit~797272
to be an X-ray flaring star from new {\em XMM-Newton} data.
Finally, Mayrit~797272 forms a visual binary ($\rho \approx$ 10.5\,arcsec,
$\theta \approx$ 5\,deg) with a slightly brighter star that does not seem to
belong to the $\sigma$~Orionis cluster.

We have derived the most probable spectral type of Mayrit~797272 in the range
M1--3V from the ALFOSC spectrum shown in Fig.~\ref{spAtoG}.
Our determination is consistent with previous analyses.
The H$\alpha$ line in our spectrum of Mayrit~797272 is sharp and with an
equivalent width similar to that one measured by Wolk (1996).
The broad emission detected by Caballero (2006) was possibly due to the low
signal-to-noise ratio and background contamination by a nearby fibre (he used a
multifibre spectrograph).
The chromospheric H$\alpha$ emission and the absence of flux excess at $\lambda
\lesssim$ 8\,$\mu$m suggest that Mayrit~797272 lacks a disc, which would make
the star to rotate slower.
The X-ray variability is easier to explain for a very young, fast-rotating,
disc-free, almost completely-convective object like Mayrit~797272 (G\"udel
et~al. 1995; Feigelson et~al. 2002; Stelzer et~al. 2004; Robrade
\& Schmitt~2005).

\subsubsection{Mayrit~102101AB} 

Mayrit~102101AB ([W96]~rJ053851--0236) was, like Mayrit~797272, first detected
as a {\em ROSAT} X-ray source by Wolk (1996).
He derived a K5 spectral type and measured EW(H$\alpha$) $\approx$ --86\,{\AA}
in a faint spectrum where he did not detect the Li~{\sc i} line.
The star has later been detected with {\em XMM-Newton} and {\em Chandra} by
Franciosini et~al. (2006), Caballero (2007b), and Skinner et~al. (2008).
Using IRAC and MIPS/{\em Spitzer} data, Hern\'andez et~al. (2007) claimed that
Mayrit~102101AB harbours an ``evolved disc'' (i.e. the star exhibits smaller
IRAC excesses than optically thick disc systems).
Remarkably, Mayrit~102101AB is one of the very few low-mass, spectroscopic
binaries (SB2) in the $\sigma$~Orionis cluster whose orbital parameters have
been derived yet ($P$ = 8.72$\pm$0.02\,d; Sacco et~al. 2008 -- they also derived
a later spectral type, K9.5, and found lithium in both spectra). 

Our observations confirm and contradict some of the observables above. 
First, using the same technique as for Mayrit~797272, we derived an M3$\pm$1
spectral type for Mayrit~102101AB, which matches better the spectro-photometric
cluster sequence than previous determinations.
The H$\alpha$ emission in our ALFOSC spectrum (EW(H$\alpha$) =
--32$\pm$2\,{\AA}) is not so strong as measured by Wolk (1996), although still
satisfying the accreting empirical criterion of Barrado y Navascu\'es \&
Mart\'{\i}n (2003 -- the White \& Basri [2003]'s criterion cannot be applied
because of our poor wavelength resolution).
The strong (variable?) H$\alpha$ emission might be originated by material
interchange between the two binary components or accretion from a disc.
As already noticed by Hern\'andez et~al. (2007), the IRAC $[3.8] - [8.0]$ colour
of Mayrit~102101AB is only of 0.27$\pm$0.05\,mag.
Furthermore, the apparent flux excess at the MIPS $[24]$ passband is due to an
overdensity in a large intracluster dust cloud, and not to the star itself
(Sect.~\ref{sigOriIRSs}).
This overdensity may also be responsible of the slight excess at $[8.0]$.
The close separation between Mayrit~102101A and~B, of $(a_1 + a_2) \sin{i}$ =
11.6$\pm$0.3\,$R_\odot$ (i.e. a few objects radii\footnote{The objects are
still contracting and have radii much larger than measured for field dwarf of
the same spectral type.}; Sacco et~al. 2008), prevents the formation of inner
disc(s) surrounding one (or both) components.
From our point of view, the origin of the X-ray and H$\alpha$ emissions is
better explained by the mutual interaction of the two binary components rather
than by the presence of an ``evolved disc''.
In any case, we do not completely discard the presence of a cool,
{\em circun-binary} disc at several tens solar~radii.

\subsubsection{Mayrit~264077} 

Mayrit~264077 (S\,Ori~J053902.1--023501) is an object at the stellar/substellar
boundary in $\sigma$~Orionis, with a most probable mass of 0.061\,$M_\odot$
(Caballero et~al. 2004, 2007).
It had, by far, the reddest near-infrared colours among the objects in the
Caballero et~al. (2007)'s survey ($J-K_{\rm s}$ = 1.84$\pm$0.05\,mag [2MASS],
$[3.6] - [8.0]$ = 1.70$\pm$0.06\,mag) and fell in a differentiated location in
the $[3.6] - [8.0]$ vs. $J-K_{\rm s}$ colour-colour diagram.
Simultaneously, Hern\'andez et~al. (2007) measured $[24]$ = 6.45$\pm$0.03\,mag
(consistent with our photometry) and sorted Mayrit~264077 as a class~II object. 
There was no spectroscopic information on it.

The spectral energy distribution in Fig.~\ref{sedAtoI} complements
that one shown in Caballero et~al. (2007).
The presence of a circumstellar disc is obvious only from photometric data;
the ALFOSC spectrum powerfully reinforces the disc existence.
We determined an M3$\pm$1 spectral type for Mayrit~264077, identical to that of
Mayrit~102101AB.
However, the H$\alpha$ emission was stronger: EW(H$\alpha$) =
--72$\pm$4\,{\AA}. 
Besides, we also found [O~{\sc i}]~$\lambda$6300.3\,{\AA} and [S~{\sc
ii}]~$\lambda\lambda$6716.4,6730.8\,{\AA} in forbidden emission (the [N~{\sc
ii}]~$\lambda$6583.4\,{\AA} is blended with the strong H$\alpha$~line).

It is worthy of notice that the $I$-band light curve of Mayrit~264077 displayed
short-, mid-, and long-scale variations in the original data of Caballero et~al.
(2004).
Since it was one of the brightest targets in their sample and its photometry
was, therefore, close to the non-linear limit of the detector, Mayrit~264077 was
conservatively discarded from the detailed analysis.
However, coming back to the original light curve, it is quite possible that
those variations were intrinsic to the object and not due to seeing variations
and non-linear effects.
As shown by Caballero et~al. (2006), T~Tauri-analog brown dwarfs in
$\sigma$~Orionis with strong H$\alpha$ emission, forbidden lines, and
near-infrared flux excess also display large amplitude photometric variations at
all considered time scales.

Mayrit~264077 might actually be a very low-mass T~Tauri star with a large
extinction bluewards of $\sim$1.5\,$\mu$m accompanied by a flux excess redwards
of it (i.e. the central object is a star brighter than $J$ =
14.44$\pm$0.04\,mag, and not a brown dwarf).

\subsubsection{Mayrit~358154} 

Mayrit~358154 (S\,Ori~J053855.4--024121) was another class~II, brown dwarf
candidate without spectroscopy in the works by Caballero et~al. (2007) and
Hern\'andez et~al. (2007).
It went unnoticed during the optical/near-infrared survey in Caballero et~al.
(2004) because it fell in the glare of the 3.8\,mag-$V$ multiple star
$\sigma$~Ori.
Since Mayrit~358154 is 1.2\,mag fainter than Mayrit~264077, we expect the
first object to be a {\em bona fide} brown dwarf, even accounting for a possible
edge-on disc. 
To date, there is only one $\sigma$~Orionis {\em star} fainter than $J \sim$
14.5\,mag, and it is the source of a well, long-time known Herbig-Haro object
(\object{HH~446} [Mayrit~633105] -- Reipurth et~al. 1998; Andrews et~al.~2004).
However, Mayrit~358154 is even fainter in $J$ than the Herbig-Haro star.
Besides, Mayrit~358154 is at 8.8\,arcsec to the north of a brighter photometric
cluster member candidate from Sherry et~al. (2004).
The hypothetical primary was not, however, in the comprehensive cluster
catalogue by Caballero~(2008c).

The ALFOSC spectrum and the spectral energy distribution strenghten the evidence
of Mayrit~358154 having a circum(sub)stellar disc. 
The flux excess at the [8.0] passband is clear, although the spectral energy
distribution seems to drop again at redder wavelengths (our upper limit at the
[24] passband is still restricting).
The distribution has a peculiar feature at the $H$ band (Mayrit~633105 should be
brighter at 1.6\,$\mu$m) whose true origin we fail to ascertain; the feature is
significative and cannot be explained by photometric error.
The H$\alpha$ emission is even larger than for the previous objects
(EW(H$\alpha$) = --190$\pm$20\,\AA) and we could identify the [O~{\sc i}],
[S~{\sc ii}], and [N~{\sc ii}] forbidden emission lines in the studied
wavelength interval. 
The derived spectral type, M5$\pm$1, is consistent with Mayrit~358154 being a
high-mass brown dwarf in $\sigma$~Orionis (the earlier brown dwarf
candidates known to date in the cluster have M5.5 spectral types).
Although it lies on the substellar domain, Mayrit~358154 displays most of the
phenomena observed in classical T~Tauri stars.

\subsection{Background galaxies}

UCM0537--0227 was the faintest new $\sigma$~Orionis photometric member candidate
in the Mayrit catalogue of Caballero (2008c).
He gave the DENIS $i$ and 2MASS $JHK_{\rm s}$ magnitudes for the object, from
where one could derive a mass in the middle of the brown dwarf domain, {\em if}
it belonged to the cluster. 
He also noticed that UCM0537--0227 possibly has an extended, non-stellar point
spread function. 
Its extremely red 2MASS ($J-K_{\rm s}$ = 2.34$\pm$0.11\,mag) and IRAC/{\em
Spitzer} colours measured by Caballero et~al. (2008) made us to obtain
spectroscopy of it.  
The low-resolution optical spectrum (top panel of Fig.~\ref{spHandI}) and
the spectral energy distribution of UCM0537--0227 correspond to an
emission-line, star-forming galaxy at a spectroscopic redshift z$_\mathrm{sp}$ =
$0.1009 \pm 0.0002$ with strong PAH features and a moderate star formation
rate.

UCM0536--0239 ([HHM2007]~668) and its odd spectral energy distribution from
the $I$ band to [8.0] was first identified by Caballero (2006). 
Just afterwards, Hern\'andez et~al. (2007) measured its [24]-band magnitude and
presented it as one of the very few class~I object {\em candidates} in
$\sigma$~Orionis. 
Caballero (2007b) noticed that UCM0536--0239 (incorrectly named
``Mayrit~111208'' in his work) had quite blue $I-J$ and extremely red $J-K_{\rm
s}$ colours for its faint magnitude ($J$ = 16.14$\pm$0.10\,mag) and found
it to be an X-ray emitter detected with the ACIS and HRC-I instruments onboard
{\em Chandra}. 
The (sub)stellar nature of UCM0536--0239 was put into question by
Caballero (2008c): he measured a possible extended point spread function in
public optical images.
Finally, Skinner et~al. (2008) measured the mean photon energy of
UCM0536--0239 at $\overline{E}$ = 3.40\,keV from ACIS/{\em Chandra} data, which
is a quite high value for a normal star, but typical of active galactic
nuclei.
The ALFOSC spectrum (bottom panel of Fig.~\ref{spHandI}) and the spectral
energy distribution of UCM0536--0239 favour its classification as a Type~1
obscured quasi-stellar object at z$_\mathrm{sp}$ = $0.2362 \pm 0.0005$.

\section{Summary}
\label{summary}

We obtained, compiled, and analyzed low-resolution spectroscopy (R $\sim$ 500),
taken with ALFOSC at the Nordic Optical Telescope, and spectral energy 
distributions, covering from the optical $V$ band to the mid-infrared MIPS/{\em
Spitzer} [24] band, of nine selected sources towards the young $\sigma$~Orionis
cluster. 
They covered a very wide range in brightness, that was expected to be translated
into a wide range in mass and spectral type.
Of the nine optical spectra presented, six correspond to objects without
previous spectroscopic observations.

Two of the targets in our sample were identified as low redshift (z $\approx$
0.10--0.24) galaxies with prominent emission-lines in the optical and large
fluxes at 8--24\,$\mu$m due to an active galactic nucleus (UCM0536--0239) and a
burst of star formation (UCM0537--0227).	
Another one seemed to be a solar-like star in the field.
Other five objects were classified as bona-fide cluster members with
spectral types in the interval A2--6 to M4--6.
We failed to detect spectroscopic features of youth in the remaining solar-like,
cluster member candidate (Mayrit~968292).

Four of the five confirmed $\sigma$~Orionis members display peculiar
features in their optical spectra and energy distributions:

\begin{itemize}
	\item Mayrit~459340 is an A2--6-type Herbig Ae/Be star with strong
	H$\alpha$ emission [EW(H$\alpha$) = --10$\pm$1\,\AA], flux excess at the
	MIPS/{\em Spitzer} [24] band, and a probable scatterer edge-on
	disc, in analogy to UX~Ori.
	\item Mayrit~102101AB is a young, low-mass, spectroscopic binary whose
	H$\alpha$ emission could be variable and originated in the mass transfer
	between A and~B.  
	\item Mayrit~264077 is an accretor at the substellar boundary with a
	developed disc and a flux excess that extends longwards of 24\,$\mu$m.
	\item Mayrit~358154 remains as the second faintest
	$\sigma$~Orionis member with spectroscopic features of youth and DENIS
	and 2MASS identification.
	Because of its strong, probably broad, H$\alpha$ emission (of up to
	about --200\,\AA), [O~{\sc i}], [S~{\sc ii}], and [N~{\sc ii}] forbidden
	lines in emission, M5$\pm$1 spectral type, flux excess at the four
	IRAC/{\em Spitzer} channels, and faintness ($i$ = 18.1$\pm$0.2\,mag), we
	classified it as a ``classical T~Tauri brown dwarf''.
\end{itemize}

The fifth cluster member, Mayrit~797272, is an X-ray flaring star
without a disc.
Besides, we also present three previuosly unknown mid-infrared sources  
without optical/near-infrared counterpart.
The $\sigma$~Orionis cluster still hides many astonishing surprises; simple
low-resolution spectroscopy and analysis of spectral energy distributions are
suitable tools for finding them.

\begin{acknowledgements}

We thank the anonymous referee, G.~Barro Calvo, M.~Rego, V.~Villar, and 
J.~Zamorano for helpful comments.  
The data presented here have been taken using ALFOSC, which is owned by the
Instituto de Astrof\'{\i}sica de Andaluc\'{\i}a (IAA) and operated at the Nordic
Optical Telescope under agreement between IAA and the NBIfAFG of the
Astronomical Observatory of Copenhagen.
Partial financial support was provided by the Universidad Complutense de Madrid,
the Spanish Virtual Observatory, the Spanish Ministerio Educaci\'on y Ciencia,
and the European Social Fund under grants AyA2005--02750, AyA2005--04286,
AyA2005--24102--E, and AyA2007--67458 of the Programa Nacional de
Astronom\'{\i}a y Astrof\'{\i}sica and by the Comunidad Aut\'onoma de Madrid
under PRICIT project S--0505/ESP--0237 (AstroCAM).  
\end{acknowledgements}

\appendix

\section{The new infrared sources $\sigma$~Ori~IRS2, 3A, and~3B}
\label{sigOriIRSs}

There exist dust clouds at relatively short separations from the Trapezium-like
system $\sigma$~Ori~AF--B. 
First, van~Loon \& Oliveira (2003) discovered \object{$\sigma$~Ori~IRS1}, a
mid-infrared source at just $\sim$3\,arcsec to the OB triple stars, that they
associated to a dust cloud. 
It is originated by the photoevaporation of the outer layers of a binary
low-mass star that lies on its densest part (Sanz-Forcada et~al. 2004; Caballero
2005, 2007a, 2007b; Turner et~al. 2008; Skinner et~al. 2008; Bouy et~al. 2008).
Next, Caballero et~al. (2008) discovered a ``cloud or shell of cool gas
surrounding [...] $\sigma$~Ori''.
Although the cloud is obvious in the MIPS/{\em Spitzer} [24]-band image
(right window in Fig.~\ref{im_all}; see also Figure~2 in Hern\'andez 
et~al. 2007), it has received no attention in the literature.

The brightest, densest knot in the cloud falls at a couple arcseconds to our
spectroscopic binary Mayrit~102101AB, whose [24]-band photometry was not
properly extracted because of the knot.
We called it $\sigma$~Ori~IRS2, following the nomenclature introduced by
van~Loon \& Oliveira (2003) for unidentified mid-infrared sources without
(known) optical/near-infrared counterpart.
There are other two additional mid-infrared sources associated to a longitudinal
arrangement in the cloud, to the northeast of $\sigma$~Ori~AF--B.
Since they seem to form part of the same structure, we called them
$\sigma$~Ori~IRS3A and~3B.

Approximate central coordinates (J2000) of the new infrared sources are
05~38~51.0 --02~36~24 (\object{$\sigma$~Ori~IRS2}),
05~38~47.6 --02~35~14 (\object{$\sigma$~Ori~IRS3A}), and
05~38~58.7 --02~35~30 (\object{$\sigma$~Ori~IRS3B}).
The origin of the dust cloud that harbours $\sigma$~Ori~IRS2, 3A, and~3B is yet
unknown.

\end{document}